# Formal Checking of Multiple Firewalls

**Nihel Ben Youssef Ben Souayeh[1] and Adel Bouhoula[2]**

[1] higher School of Communication of Tunis (Sup'Com)
University of Carthage, Tunisia
**nihel.byoussef@supcom.rnu.tn**

[2] higher School of Communication of Tunis (Sup'Com)
University of Carthage, Tunisia
**adel.bouhoula@supcom.rnu.tn**

**Abstract**
When enterprises deploy multiple firewalls, a packet may be examined by different sets of firewalls. It has been observed that the resulting complex firewall network is highly error prone and causes serious security holes. Hence, automated solutions are needed in order to check its correctness. In this paper, we propose a formal and automatic method for checking whether multiple firewalls react correctly with respect to a security policy given in a high level declarative language. When errors are detected, some useful feedback is returned in order to correct the firewall configurations. Furthermore, we propose a priority-based approach to ensure that no incoherencies exist within the security policy. We show that our method is both correct and complete. Finally, it has been implemented in a prototype of verifier based on a satisfiability solver modulo theories. Experiment conducted on relevant case studies demonstrates the efficiency of our approach.
.

**Keywords:** *network security, distributed firewall configuration, formal verification, SMT solver.*

## 1. Introduction

Firewalls are the most widely adopted technology for protecting private networks. Placed, generally, at the point of entry between public network and private network zones, a firewall ensures the access control of the forwarding traffic. However, according to the study undertaken by Wool [15], most firewalls in Internet are plagued with policy errors. The main firewall configuration constraint is that the filtering rules of a firewall configuration FC file are treated in the order in which they are read in the configuration file, in a switch-case fashion. For instance, if two filtering rules associate different actions to the same flow type, then only rule with the lower order is really applied. This is in contrast with the security policy SP, which is a set of rules considered without order. In this case, the action taken, for the flow under consideration, can be the one of the non-executed rule. The following example illustrates how easily firewall mis-configurations can happen:

Table 1: Firewall Configuration Error

|    | *src_adr*     | dst adr | protocol | dst_port | action |
|----|---------------|---------|----------|----------|--------|
| r1 | 214.0.0.0/8   | *       | tcp      | *        | accept |
| r2 | 214.65.0.0/16 | *       | tcp      | 445      | deny   |

The second rule is configured to deny all the outbound traffic to a known backdoor TCP port for the sasser worm Which is conform to a specific SP . But even if this rule is correct by itself, the firewall will accept this flow type because it matches the rule before. In this case, r1 shadows r2 and leaves the hole wide open. A correct configuration according to this specific SP could be a swap of the two rules.

As shown by Chapman [18], safely configuring firewall rules has never been an easy task. Since, firewall configurations are low-level files, subject to special configuration constraints in order to ensure an efficient real time processing by specific devices. Whereas, the security policy SP , used to express global security requirements, is Generally specified in high-level declarative language easy to understand. Hence, this makes verifying the conformance of a firewall configuration FC to a security policy SP a daunting task. Particularly, when it is to analyze the impact of the inter-actions of a large number of rules on the behavior of a firewall. Moreover, when large enterprise deploy multiple firewalls to manage internal traffic between private zones due to the growing number of internal attacks, a packet from the same source to the same destination may be examined by different sets of firewalls. It is so even more difficult to check whether all these sets of firewalls satisfy the end-to-end security policies of the enterprise.

Several methods have been proposed [14], [2], [4], [1], [3], [24], [21] for the detection of inter-rule conflicts in FC. These work are limited to the problem conflict avoidance, and do not consider the more general problem of verifying whether a firewall reacts correctly with respect to a given

SP. Solutions are studied in [11], [7], [16], [23], [13] for the analysis of firewalls' behavior. These methods require some final user interactions by sending queries through a verification tool. Such manual solutions can be tedious when checking discrepancies with respect to complicated security requirements. In [5],[12] and [10] the authors address the problem of automatic verification by providing automatic translation tool of the security requirements (SP), specified in a high level language, into a set of ordered filtering rules (i.e. a FC). Therefore, these methods can handle the whole problem of conformance of FC to SP, but the validity of the compilation itself has not been proved. In particular, the FC rules obtained may be in conflict. In our previous work [19], we proposed to verify the correctness of a single firewall configuration according to a given SP. In this paper, we consider the more general and complex case by proposing an automatic method for checking whether a distributed firewall is well configured according to a global security policy, given in an expressive enough declarative language. Furthermore, the proposed method ensures conflicts avoidance within the SP that we aim to establish and returns key elements for the correction of flawed firewall configurations. Our method has been implemented as a prototype which can be used either in order to validate an existing distributed FC with respect to a given SP or downstream of a compiler of SP. It can also be used in order to assist the updates of a distributed FC, since some conflicts may be created by the addition or deletion of filtering rules. The work of Liu and Gouda [17] is similar to ours in spirit. However, their solution is applied for one specific security property and considered exactly one possible path from a source to a destination zone. As shown above, it seems necessary to assume that all paths are topologically possible for ensuring the correctness of proposed algorithms. Besides that, routing is designed to be adaptive to link failures and heavy load. In addition, it is relatively easy to inject routing messages [22]. In other hand, strengths compared to their work consist on : First, proving the correctness and the completeness of our method and second, giving key elements with high level granularity to help the correction of firewall mis-configurations which should be the main and the concrete target of such study.

## 2. Security Policy

### 2.1 Formal Specification

A security policy (SP) is a finite set of security directives defining whether packets are accepted or denied: $SP = \{sd_i \Rightarrow A_i \,|[e_i\,]\,|\,1 \leq i \leq n\}$. Each security directive can be simple or complex. A simple directive $\{sd_i \Rightarrow A_i\}$ describes whether some traffic destined to one or more services that are required by one or more sources and given by one or more destinations (as described by the condition $sd_i$) must be accepted or refused (according to $A_i \in \{accept, deny\}$). A complex directive $\{sd_i \Rightarrow A_i \,|e_i\}$ is basically a simple directive with some additional exceptions defined in $e_i$. In our previous work [19], we consider only one exception in our verification process. The following examples are simple and complex directives.

- The sub zone LAN'_A of LAN_A has not the right to access to the FTP server located in LAN_B.
- The zone LAN_A has not the right to access to the zone LAN_B. However, the machine A1 in LAN_A can access to LAN_B and the sub zone LAN"_A has the right to access to the FTP server located in LAN_B.
- The machine A2 in LAN_A has not the right to access to the sub zone LAN'_B of LAN_B.

We note that LAN'_A and LAN"_A have a set M of common machines. As follows, a formal definition of the above security policy:

$$\begin{pmatrix} sd_1 & \Rightarrow & deny \\ sd_2 & \Rightarrow & deny & | & e_{21} & , & e_{22} \\ sd_3 & \Rightarrow & deny \end{pmatrix}$$

Fig 1. Formal Specification of a Security Policy.

Let we consider definition domain of SP, partitioned the into $dom(SP) = \bigcup_{A \in \{accept, deny\}} SP_A$. Each set $SP_A$ is composed by a set of domains $SP_{Ai}$ of security directives concerning a specific flow from a source sr to a destination dt: $SP_A = \{SP_{Ai}\,(sr, dt)\,|\,1 \leq i \leq l\}$. Each set $SP_{Ai}$ represents either the domain of a simple directive, if the action of the corresponding condition is A, or the domain of a complex directive's exception, if its action is Accept or else the difference between the domains of the condition and the exceptions of a complex directive if its main action is A. Formally, $SP_{Ai} = \{dom(sd_i \Rightarrow A_i) \setminus [dom(e_i))]\|dom(e_i \Rightarrow A); 1 \leq j \leq n\}$. In this case, $SP_{accept} = dom(e_{21}) \cup dom(e_{22})$ and $SP_{deny} = dom(sd_1) \cup dom(sd_2) \setminus dom(e_{21}, e_{22}) \cup dom(sd_3)$. For the next section, let we consider $SD_{Ai}$ the security element whose domain is $SP_{Ai}$. And let $SD_A$ be the set of such elements.

### 2.2 Fixing Security Policy Incoherencies

First, let we consider $Elts_A$ as the set of individual elements in SP labeled by the action A. Formally, $Elts_A = \{sd_i \Rightarrow A_i \,|e_i\,\}$. For example, in the SP defined in previous

$$\text{init}_n \quad \frac{}{F_n, \emptyset, dom(SP_{a_i})}$$

$$\text{recurcall}_{d_n} \quad \frac{((r \Rightarrow d), F_n), D, D_{d_n}}{F_n, D \cup dom(r), D_{d_n}} \quad \text{if } dom(r) \setminus D \cap dom(SP_{a_i}) = \emptyset$$

$$\text{recurcall}_{a_n} \quad \frac{((r \Rightarrow a), F_n), D, D_{d_n}}{F_n, D \cup dom(r), D'_{d_n}} \quad \text{where } \begin{cases} D'_{d_n} = D_{d_n} \setminus (dom(r) \setminus D) & \text{if } def(F_n) = \text{deny} \\ D'_{d_n} = D_{d_n} & \text{otherwise} \end{cases}$$

$$\text{success}_n \quad \frac{\emptyset, D, D_{d_n}}{\text{success}} \quad \text{if } def(F_n) = \text{deny and } D_{d_n} = \emptyset$$

$$\text{failure}_n \quad \frac{F_n, D, D_{d_n}}{\text{fail}} \quad \text{if no other rule applies}$$

Fig 2. Inference System for a SPai

section, $Elts_{deny} = \{sd1, sd2, sd3\}$ and $Elts_{accept} = \{e21, e22\}$. Let *Before_Ai* be the set of elements in $Elts_A$ that should have higher priority than that of $elt_{Ai}$. Once SP specified in expressive enough language, our goal is to certify that no contradictions exist within security directives. To verify SP coherent, we should determine whether $SP_{accept} \cap SP_{deny} = \phi$. In negative cases, this means that there exists at least a couple of elements ($elt_{Ai}$, $elt^c{}_{Ai}$) that impose each contradictory actions for common packets involved in their effective domains. Let *Conflict* be the set of such couples. In the security policy given as example in section 2.1, we can note that the first two directives are in conflict. Particularly, ($elt_{deny1}$, $elt_{accept2}$). Indeed, $sd_1$ indicates that the sub zone LAN'A has not the right to access FTP server. Whereas, the set M of machines common to LAN'A and LAN"_A should be authorized according to $e_{22}$. Once our method outputs these results, the administrator should define which of the elements should be considered by priority (ie. The common machines M have or not the right to access FTP server). For instance, if, in our example, the administrator judges that this access should be prohibited then $elt_{deny1}$ has higher priority than $elt_{accept2}$. The set *Before_Ai* is so expressed as follows:

| | |
|---|---|
| $Before\_accept_1$ :: | $\{\}$ |
| $Before\_accept_2$ :: | $\{sd_1\}$ |
| $Before\_deny_1$ :: | $\{\}$ |
| $Before\_deny_2$ :: | $\{e_{21}, e_{22}\}$ |
| $Before\_deny_3$ :: | $\{\}$ |

Thus, each element $SP_{Ai}$ of $SP_A$ is newly defined as follows, ***$SP_{Ai} = dom(elt_{Ai}) \setminus dom(Before\_Ai)$***. In our case, $SP_{accept} = dom(e21) \cup dom(e22) \setminus dom(sd1)$ and $SP_{deny} = dom(sd1) \cup dom(sd2) \setminus (dom(e21) \cup dom(e22) \setminus dom(sd1)) \cup dom(sd3)$.

## 3. Conformance Properties

The main goal of this work consists of checking whether a distributed FC is conform to a given SP. In this section, we define formally this notion. We consider a finite domain P containing all the headers of packets possibly incoming to or outgoing from a network. A simple firewall configuration (Fn) is a finite sequence of filtering rules of the form $Fn = (ri \Rightarrow Ai)_{0 \leq i < m}$. Each precondition ri of a rule defines a filter for packets of P. The structure of ri is described later in Section 5. Until then, we just consider a function dom mapping each ri into the subset of P of filtered packets. Each right member Ai of a rule of FC is an action defining the behavior of the firewall on filtered packets: $Ai \in \{accept, deny\}$. If no filtering rule ri can be considered for a specific packet, the default firewall policy will be applied : $def(Fn) \in \{accept, deny\}$. This model describes a generic form of FC which are used by most firewall products such as CISCO, Access Control List, IPTABLES, IPCHAINS and Check Point Firewall...

A Path(sr, dt) is an ordered set of firewalls through which the traffic flow (sr → dt) could go across : Path(sr, dt) = (Fi | 1 ≤ i ≤ N). Let [[Path(sr, dt)]] be the set of all possible paths from sr to dt.

A distributed FC is conform to a SP if the action defined by SP for each packet p concerning a traffic from sr to dt is really undertaken by the distributed firewall. Precisely, we distinguish two cases:
- For each positive security rule $SP_{Ai}$, p should be accepted whatever the path to cross. This implies that p should be allowed by each firewall Fn belonging to each path.
- For each restrictive security rule $SP_{di}$, p should be denied whatever the path to cross. This implies that p should be denied by at least one firewall Fn belonging to each path.

$$\text{init}_n \quad \frac{}{F_n, \emptyset, D_{a_n}} \quad with \quad \begin{cases} D_{a_n} = \emptyset & if\ def(F_n) = deny \\ D_{a_n} = D_{a_{n-1}} & Otherwise \end{cases}$$

$$\text{recurcall}_{a_n} \quad \frac{((r \Rightarrow a), F_n), D, D_{a_n}}{F_n, D \cup dom(r), D'_{a_n}} \quad with \quad D'_{a_n} = \begin{cases} D_{a_n} \cup (dom(r) \setminus D) & if\ def(F_n) = deny \\ D_{a_n} & Otherwise \end{cases}$$

$$\text{recurcall}_{d_n} \quad \frac{((r \Rightarrow d), F_n), D, D_{a_n}}{F_n, D \cup dom(r), D'_{a_n}} \quad with \quad D'_{a_n} = \begin{cases} D_{a_n} \setminus (dom(r) \setminus D) & if\ def(F_n) = accept \\ D_{a_n} & Otherwise \end{cases}$$

$$\text{success}_n \quad \frac{\emptyset, D, D_{a_n}}{success} \quad with \quad \begin{cases} Da_n = \emptyset & if\ def(F_n) = accept \\ or & \\ Da_{n-1} \cap Da_n = \emptyset & if\ def(F_n) = deny \\ else & \\ success_{n+1}\ with\ Da_n = Da_{n-1} \cap Da_n & if\ n < |\mathcal{P}a| \end{cases}$$

$$\text{failure}_n \quad \frac{F_n, D, D_{a_n}}{fail} \quad if\ no\ other\ rule\ applies$$

Fig 3. Inference System for a SPdi

*Definition 1 (conformance property for Spai)* : A distributed F C is conform to SPai (sr, dt) iff $\forall p \in dom(SPai\ (sr, dt))$, $\forall Pa \in [[Path(sr, dt)]]$ and $\forall Fn \in Pa$, AFn (p) = accept.

*Definition 2 (conformance property for Spdi)* : A distributed F C is conform to SPdi (sr, dt) iff $\forall p \in dom(SPdi\ (sr, dt))$, $\forall Pa \in [[Path(sr, dt)]]$, $\exists Fn \in Pa$, AFn (p) = deny.

AFn (p) represents the action undertaken by the firewall Fn for a packet p. It is defined as follows: when def (Fn ) = deny, if there exists a rule ri ⇒ a in Fn such that $p \in dom(ri) \setminus \cup_{j<i} dom(rj)$, AFn (p) = accept otherwise, AFn (p) = deny.

Let $Acc_n$ and $Den_n$ be respectivelly the set of accepted and denied packets by Fn . $Acc_n$ is defined in this case as follows: Accn = $\cup_i (dom(ri) \setminus \cup_{j<i} dom(rj))$ with ri ⇒ a.

By analogy, when def (Fn ) = accept, if there exists a rule ri ⇒ d in Fn such that $p \in dom(ri) \setminus \cup_{j<i} dom(rj)$, AFn (p) = *deny*. Otherwise, AFn (p) = *accept*. Therefore, $Den_n$ is defined as follows: $Den_n$ =$\cup_i (dom(ri) \setminus \cup_{j<i} dom(rj))$ with ri ⇒ d. And in each case, $Acc_n$ and $Den_n$ are complementary.

## 4. Inference Systems

We propose, in this section, necessary and sufficient conditions for the verification of the conformance property of a distributed FC to a SP. The conditions are presented mainly as inference systems shown in Figure 2 and Figure 3. The first inference system in Figure 2 concerns each firewall Fn in all paths belonging to [[Path(sr, dt)]], where sr and dst represent the source and the destination fields of a positive security rule SPai . The rules of the system in Figure 2 apply to triples (Fn , D, $D_{dn}$ ) whose first component Fn is a sequence of filtering rules and whose second and third components ,respectively D and $D_{dn}$ are subsets of P. D represents the accumulation of the sets of packets filtered by the rules of Fn processed so far. $D_{dn}$ represents the sets of packets considered by SPai and not filtered by the rules of Fn labeled by positive actions.

We write C |-SP C' : C' is obtained from C by application of one of the inference rules of Figure 2 and Figure 3 (note that C' may be a triple as above or one of *success* or *fail*) and we denote by |- SP * the reflexive and transitive closure of |- SP.

*recurcall*$_{an}$ and *recurcall*$_{dn}$ are the main inference rules. For the inference system in Figure 1, *recurcall*$_{dn}$ deals with the first filtering rule r ⇒ d of Fn given in the couple. The condition for the application of *recurcall*$_{dn}$ is that the set of packets dom(r) filtered by this rule and not handled by the previous rules (i.e. not in D) would not intersect the domain of SPai . The inference rule *recurcall*$_{an}$ deals with the first filtering rule r ⇒ a of Fn given in the couple. The condition for its application is that the default firewall policy is deny. It results in excluding the effective part of the rule r from the set $D_{dn}$ . Hence, successful repeated applications of *recurcall*$_{dn}$ and *recurcall*$_{an}$ ensure that the Fn under consideration is conform to SPai. The *success*$_n$ rule is applied under two conditions. First, *recurcall*$_{dn}$ must have been used successfully until all filtering rules have been processed (in this case the first component Fn of the triple is empty). Second, the set $D_{dn}$ should be empty if the default firewall policy is *deny*. This latter condition ensures that all the packets accepted by the security rule SPai are also handled by the firewall configuration. There are two cases for the application of f*ailure*$_n$. In the first case, f*ailure*$_n$ is applied to a triple (Fn , D, $D_{dn}$ ) where Fn is not empty. It means that *recurcall*$_{dn}$ has failed on this triple and hence that the Fn is not conform to SPai . In this case, *failure*$_n$ returns the first filtering rule of Fn as an example of rule which is not correct, in order to provide help to the user for correcting the FC. In the second case, *failure_n* is applied to ($\phi$, D, $D_n$). It means that *success_n* has failed on this triple and that the $F_n$ is not conform to $SP_{ai}$. In this case, $D_{dn}$ is returned and can be used in order to identify packets accepted by the SP and not by the $F_n$.

The second inference system in Figure 3 concerns the first firewall $F_n$ for each path belonging to [[Path(sr, dt)]], where sr and dst represent the source and the destination fields of a restrictive security rule $SP_{di}$. The rules of the system in Figure 3 apply to triples ($F_n$, D, $D_{an}$) whose first component $F_n$ is a sequence of filtering rules and whose second and third components, respectively D and $D_{an}$ are subsets of P. $D_{an}$ is initialized to $\phi$ if the default policy of $F_n$ is *deny* and to $D_{an-1}$. Otherwise, relatively to the previous firewall number n − 1 belonging to the same path. For this inference system, the inference rule *recurcallan* deals with the first filtering rule r ⇒ a of $F_n$ given in the couple. The condition for the application of *recurcallan* is that the default firewall policy is *deny*. It results in accumulating the effective part of the rule r to the set $D_{an}$. The next inference rule, *recurcalldn* deals with the first filtering rule r ⇒ d of $F_n$ given in the couple. The condition for the application of *recurcalldn* is that the default firewall policy is *accept*. It results in excluding the effective part of the rule r from the set $D_{an}$. The *success_n* rule is applied when, first, *recurcalldn* and *recurcallan* have been used successfully until all filtering rules have been processed (in this case the first component $F_n$ of the triple is empty). And second, at least one the following conditions holds:

- The set $D_{an}$ is empty if the default firewall policy is *accept*. This condition ensures that the packets considered by $SP_{di}$ but allowed by the (n − 1) previous firewalls of the same path are totally denied by $F_n$.

- The intersection of the sets $D_{an}$ and $D_{an-1}$ is empty if the default policy of $F_n$ is *deny*. This condition guarantees that the packets considered by $SP_{di}$ but allowed by the (n − 1) previous firewalls of the same path are not allowed by $F_n$.

The *follow_n* rule applies if the conditions of the *success_n* rule are not satisfied and the firewall $F_n$ under consideration is not the last in the path Pa. Applying this rule updates the set $D_{an}$ of accepted packets passed through the n firewalls, although they should be denied according to $SP_{di}$. The application of *failure_n* is triggered when, either, n = |Pa| and (R = $D_{an}$) = $\phi$ if the default $F_n$ policy is *accept* or (R = $D_{an}$ ∩ $D_{an-1}$) = $\phi$, otherwise. The two cases mean that the set R of packets will be allowed by the chain of firewalls composing the path Pa, which dissent to $SP_{di}$. If this inference rule occurs, our tool outputs the set R indicating the path Pa under consideration to help the user to correct its configuration. Let us now prove that the inference systems presented in Figure 2 and Figure 3 are correct and complete. From now on, we assume that SP is consistent. This implies that $\forall i$, $\forall j$, $SP_{di} \cap SP_{aj} = \phi$.

Thus, the theorems below deal with generic cases for distinct security rules $SP_{di}$ and $SP_{ai}$.

**Theorem 1 (correctness):** For a $SP_{ai}$ (sr, dt), if $\forall Pa \in$ [[Path(sr, dt)]] and $\forall F_n \in$ Pa, such that ($F_n$, $\emptyset$, dom($SP_{ai}$)) |-*$_{SPai}$ success then the distributed firewall configuration FC is conform to $SP_{ai}$.

*Proof*: If for a $SP_{ai}$ (sr, dt), $\forall Pa \in$ [[Path(sr, dt)]] and $\forall F_n \in$ Pa, ($F_n$, $\emptyset$, dom($SP_{ai}$)) |- * $_{SPai}$ success then we have two cases: if def ($F_n$) = accept then for all p ∈ $Sp_{ai}$, $\forall r_i \Rightarrow d$, $p \notin$ dom($r_i$) \ $\cup_{j<i}$ dom($r_j$) through the condition of recullcalldn. Hence, $A_{Fn}$ (p) = accept. Second, if def ($F_n$) = deny, then dom($SP_{ai}$) \ $\cup_i$ (dom($r_i$) \ $\cup_{j<i}$ dom($r_j$)) = $\emptyset$. by the application of *recullcallan*. In this case, for all p ∈ dom($SP_{ai}$), there exists $r_i \Rightarrow a$ such that p ∈ dom($r_i$) \ $\cup_{j<i}$ dom($r_j$). Hence, $A_{Fn}$ (p) = accept. Therefore, the distributed FC is conform to $Sp_{ai}$.

**Theorem 2:** For a $SP_{di}$ (sr, dt), if $\forall Pa \in$ [[Path(sr, dt)]], $\exists n \in$ Pa, such that ($F_n$, $\emptyset$, $D_{an}$) |-* $_{SPdi}$ success then the distributed firewall configuration FC is conform to $SP_{di}$.

*Proof*: For a $SP_{di}$ (sr, dt), if $\forall Pa \in$ [[Path(sr, dt)]] and $\exists F_n \in$ Pa, such that ($F_n$, $\emptyset$, $D_{an}$) |-* $_{SPdi}$ success then $D_{an-1} \setminus Den_n = \emptyset$ if def ($F_n$) = accept or $D_{an-1} \cap Acc_n = \emptyset$, otherwise. This guarantees that the set of packets included in $D_{an-1}$ are totally denied by $F_n$. Moreover, $D_{an-1}$ = $D_{an-2} \setminus Den_{n-1}$ if def ($F_n$) = *accept* or $D_{an-1} = D_{an-2} \cap Acc_{n-1}$, otherwise. Thus, $D_{an-1}$ represents in the two cases, the set of packets included in $D_{an-2}$ and denied by $F_{n-1}$. With $D_{a0}$ = dom($SP_{di}$), We can easily show by induction on *n* that $D_{an-1}$ represents the set of packets belonging to dom($SP_{di}$) and not denied by any of the (n − 1) previous firewalls. It implies that dom($SP_{di}$) ⊆ $\cup_{(1 \leq i \leq n)} Den_i$. It follows that $\forall p \in$ dom($SP_{di}$), $\exists F_n$, such that $A_{Fn}$ (p) = *deny*. Hence, the distributed firewall is conform to $SP_{di}$.

**Theorem 3:** The distributed firewall configuration FC is conform to $SP_{ai}$ (sr, dt) iff $\forall Pa \in$ [[Path(sr, dt)]] and $\forall F_n \in$ Pa, ($F_n$, $\emptyset$, dom($SP_{ai}$)) |- $_{SPai}$ * success.

*Proof:* The distributed firewall configuration FC is conform to $SP_{ai}$ (sr, dt) implies that $\forall Pa \in$ [[Path(sr, dt)]], $\forall F_n \in$ Pa and $\forall p \in$ P, we have p ∈ dom($SP_{ai}$) and p ∈ $Acc_n$. It implies that dom($SP_{ai}$) \ $Acc_n = \emptyset$ if def ($F_n$) = deny. And, $\forall r_i \Rightarrow d$, dom($r_i$) \ $\cup_{j<i}$ dom($r_j$) ∩ dom($SP_{ai}$) = $\emptyset$, otherwise. Hence, successful repeated applications of

*recurcallan* and *recurcalldn* rise to $(F_n, \emptyset, dom(SP_{ai}))$ |-$_{SP_{ai}}$ * success.

**Theorem 4:** The distributed firewall configuration FC is conform to $SP_{di}$ (sr, dt) iff $\forall P_a \in [[Path(sr, dt)]] \exists F_n \in P_a$, such that $(F_n, \emptyset, D_{an})$ |-* $_{SP_{di}}$ success.

*Proof:* The distributed firewall configuration FC is conform to $SP_{di}$ (sr, dt) implies that $\forall p \in dom(SP_{ai})$, $\exists F_n \in P_a$ such that $p \in Den_n$. It implies that $dom(SP_{di}) \subseteq \cup_{(1 \leq i \leq |<pa|)} Den_i$. As shown in Theorem 2, this case is occurred when $(F_n, \emptyset, D_{an})$ |-* $_{SP_{di}}$ success is reached.

**Theorem 5:** If $(F_n, D, D_{dn})$ |-* $_{SP_{ai}}$ fail then the distributed firewall configuration F C is not conform to $SP_{ai}$.

*Proof:* Either we can apply iteratively the *recurcallan* and *recurcalldn* rules starting with $(F_n, \emptyset, SP_{ai})$, until we obtain $(\emptyset, \cup_{j<n} dom(r_j), D_{dn})$, or one application of the *recurcalldn* rule fails. In the latter case, there exists (i < n) $\Rightarrow$ d such that $dom(r_i) \setminus \cup_{j<i} dom(r_j) \cap SP_{ai} = \emptyset$. Therefore, there exists $p \in P$ such that $p \in dom(r_i) \setminus \cup_{j<i} dom(r_j)$ and $p \in SP_{ai}$. It follows that FC is not conform to the security policy SP. If $(F_n, \emptyset, SP_{ai})$ |-* $_{SP_{ai}}$ $(\emptyset, \cup_{j<n} dom(r_j), D_{dn})$ using *recurcallan* and *recurcalldn* but the application of the *successn* rule to the last triple fails, then there exists $D_{dn} = \emptyset$ if def $(F_n) = deny$. It means that $dom(SP_{ai}) \setminus Acc_n = \emptyset$. It follows that $\exists p \in P$, such that $p \in dom(SP_{ai})$ and $p \notin Acc_n$. Hence, the distributed firewall configuration FC is not conform to the security policy $Sp_{ai}$.

**Theorem 6:** If $(F_n, D, D_{an})$ |-$_{SP_{di}}$* fail then the distributed firewall configuration FC is conform to $SP_{di}$.

*Proof:* If $(F_n, D, D_{an})$ |- $_{SP_{di}}$ * fail then either $D_{an} = \emptyset$ or $D_{an} \cap D_{an-1} = \emptyset$ with n = |Pa|. The two cases occur, as shown in Theorem 2, if $dom(SP_{di}) \not\subset \cup_{(1 \leq i \leq n)} Den_i$. It follows that, $\exists p \in dom(SP_{di})$, such that, $F_n$ with $A_{F_n}(p) = deny$. Hence, the distributed firewall configuration FC is not conform to the security policy $SP_{di}$. Since the application of the inferences to $(F_n, \emptyset, dom(SP_{dai})$ and $(F_n, \emptyset, D_{an})$ of respectively the inference systems in Figure 2 and Figure 3 always terminate, and the outcome can only be success or fail, it follows immediately from Theorem 1 and Theorem 2 that if the firewall configuration FC is not conform to either $SP_{ai}$ or $SP_{di}$, then $(F_n, \emptyset, D_{ai})$ |- $_{Spai}$ * fail (completeness of failure).

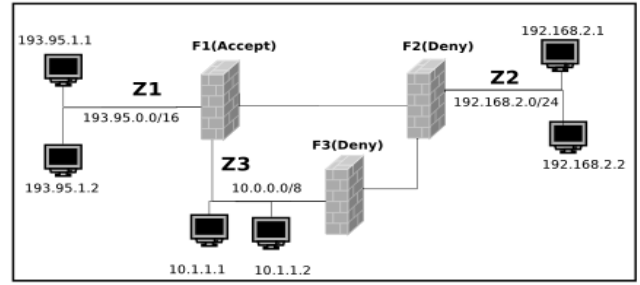

Fig 4. A Corporate Network

## 5. Automatic Verification Tool

Presenting the above Conformance Properties as satisfiability problems permits the automation of the verification of the conformance of a distributed FC to a SP. For this purpose, we have used a recent satisfiability solver modulo theories, Yices [6], in order to describe the different inputs and to automate the verification process. Yices provides different additional functions, compared to simple satisfiability solvers. These functions are based on theories like those of arrays, list structures and bit vectors. The first input of our verification tool is a set of firewalls. Each firewall is composed by a set of filtering rules. Each rule is defined by a priority order and composed of the following main fields: the source, the destination, the protocol and the port. The source and destination fields correspond to one or more machines identified by an IPv4 address and a mask coded both on 4 bytes, For example, the following expression written in Yices syntax refers to the third filtering rule concerning UDP or TCP flow, coming from the source network 10.0.0.0/8 and reaching the network 192.168.0.0/16 for a destination port belonging to the subrange [20 − 60].

*(define r :: (-> int bool))(assert (= (r 3) (and (= ips1 10) (= ipd1 192) (= ipd2 168) (>= port 20) (<= port 60) (or (= protocol tcp) (= protocol udp))))).*

In order to illustrate the proposed verification procedure, we have chosen to apply our method to a case study of a corporate network represented in Figure 4. The network is divided into three zones delineated by branches of firewalls $F_1$, $F_2$, $F_3$ whose initial configurations FC corresponds to the rules in figure 5.

| | src_adr | dst_adr | protocol | dst_port | action |
|---|---|---|---|---|---|
| F1(1) | 193.95.0.0/16 | 10.1.1.2 | 23 | tcp | accept |
| F1(2) | 193.95.0.0/16 | 10.0.0.0/8 | * | * | deny |
| F1(3) | 10.0.0.0/8 | 192.168.2.1 | 53 | udp | accept |

| | src_adr | dst_adr | protocol | dst_port | action |
|---|---|---|---|---|---|
| F2(1) | 193.95.0.0/16 | 10.1.1.2 | 23 | tcp | accept |
| F2(2) | 10.0.0.0/8 | 192.168.2.2 | 22 | tcp | accepts |
| F2(3) | 193.95.0.0/16 | 192.168.2.0/24 | * | * | accept |

| | src_adr | dst_adr | protocol | dst_port | action |
|---|---|---|---|---|---|
| F3(1) | 193.95.0.0/16 | 192.168.2.0/24 | 22 | tcp | deny |
| F3(2) | 193.95.0.0/16 | 10.1.1.2 | 23 | tcp | accept |
| F3(3) | 10.0.0.0/8 | 192.168.2.1 | 53 | udp | accept |
| F3(4) | 193.95.0.0/16 | 192.168.2.0/24 | * | * | accept |

Fig 5. Distributed Firewall Configuration

The security policy SP that should be respected contains the following directives.

*sd 1 : The zone Z1 has not the right to access to The zone Z3 .*
*sd 2 : The zone Z3 has not the right to access to the SSH server Z2.*
*sd 3 : The zone Z1 has the right to access to the TELNET server Z'3 .*
*sd 4 : The zone Z3 has the right to access to the DNS server Z".2*
*sd 5 : The zone Z1 has the right to access to The zone Z2 , except to its SSH server Z'2 .*

As defined in section 2 , [[Path(sr, dt)]] is the set of all possible paths from a source sr to a destination dt. In this case, we have :
   *[[Path(Z1 , Z3 )]] ::= {(F1 ), (F1 , F2 , F3 )}*
   *[[Path(Z1 , Z2 )]] ::= {(F1 , F2 ), (F1 , F3 , F2 )}*
   *[[Path(Z3 , Z2 )]] ::= {(F1 , F2 ), (F3 , F2 )}*

5.1 Security Policy Coherence

As previously mentioned, we should first check whether the security policy is coherent. By implementing our verification method using Yices, the satisfiability result obtained is displayed in figure 6. The outcome shows that SP is not coherent i.e. that the security directives sd1 and sd3 have contradictory actions for common packets.

```
(= (sd 1) true)
(= (sd 2) false)
   (sd 3) true)
(= (sd 4) false)
(= (sd 5) false)
(= (e 5 1) false)
```

Fig 6. Checking SP Coherence

Indeed, the zone Z1 has the right to access to the TELNET server according to sd3. Whereas, sd1 denies this access. To fix such incoherence, the administrator could fix which of the two elements has higher priority. For example, let we consider that elt_deny1 has higher priority than elt_accept1. In this case, sd1 could be replaced by a complex directive as follows:
*sd 1 : The zone Z1 has not the right to access to The zone Z3, except to its TELNET server .*

5.2 Conformance Verification of positive policies

Once ensured that SP is coherent, we proceed to the verification of the conformance of the distributed FC to each positive security rule. The first satisfiability result obtained is displayed in figure 7. The outcome shows that the distributed F C is not conform to SP . i.e. that some packets that should be accepted according to spa2 are denied by the second firewall of the first path of [[Path(Z3 , Z2 )]] which is F2. It indicates also, that no rule is accepting this type of traffic. Therefore, these packets are denied by the default firewall policy of F2 (deny). This conflict can be resolved by adding a rule at the end of the F2 configuration to deploy SPa2.

```
root@nihel:/home/nihel/Desktop/yices-1.0.21/bin# ./yices -e run.ys
sat
(= protocol udp)
(= ips1 10)
(= ips2 96)
(= ipd1 192)
(= ipd2 168)
(= ipd3 2)
(= ipd4 1)
(= port 53)
(= Path 1)
(= FwNum 2)
(= (sp' 1) false)
(= (spa 2) true)
(= (sp 3) true)
(= (spa 3) false)
(= (sp 4) false)
(= (sp' 4) false)
(= (spd 1) false)
(= (sp 1) false)
(= (spd 2) false)
(= (sp 2) false)
(= (spd 3) false)
(= (F 1 1) false)
(= (F 1 2) false)
(= (F 1 3) true)
(= (F 2 1) false)
(= (F 2 2) false)
(= (F 2 3) false)
(= (F 3 1) false)
(= (F 3 2) true)
(= (F 3 3) false)
```

Fig. 7 Automatic Verification of spa2

5.2 Conformance Verification of restrictive policies

After that the conformance property to positive policies has been established, we proceed to the verification of the distributed F C to the restrictive policies. We obtained the satisfiability result displayed in Figure 8.

```
root@nihel:/home/nihel/Desktop/yices-1.0.21/bin# ./yices -e run.ys
sat
(= protocol tcp)
(= ips1 10)
(= ips2 96)
(= ipd1 192)
(= ipd2 168)
(= ipd3 2)
(= ipd4 2)
(= port 22)
(= Path 1)
(= FwNum 2)
(= (spa 1) false)
(= (sp' 1) false)
(= (spa 2) false)
(= (sp 3) false)
(= (spa 3) false)
(= (sp 4) false)
(= (sp' 4) false)
(= (spd 1) false)
(= (sp 1) false)
(= (spd 2) true)
(= (sp 2) true)
(= (spd 3) false)
(= (F 1 1) false)
(= (F 1 2) false)
(= (F 1 3) false)
(= (F 2 1) false)
(= (F 2 2) true)
(= (F 2 3) false)
(= (F 2 4) false)
(= (F 3 1) false)
(= (F 3 2) false)
(= (F 3 3) false)
(= (F 3 4) false)
(= (F 3 5) false)
```

Fig. 8 Automatic Verification of spd2

According to this outcome, the distributed F C is not conform according to $SP_{a2}$ : There are some packets handled by $SP_{a2}$ that will be accepted by crossing the first path (F1 , F2 ) until the firewall F2 . The outcome indicates also that the second filtering rule of F2 is accepting some packets previously allowed by F1 , which is in conflict with the requirements of SPd2 . Indeed, the rule F2 (2) implements totally the condition of SPd2 but the action considered is accept. This conflict can be resolved by changing the later by deny. We note that YICES ensures the conformance of $SP_{a1}$, $SP_{d1}$ ,and $SP_{d3}$. Figure 9 presents a correct and complete distributed configuration according to the defined SP .

|        | src_adr       | dst_adr       | protocol | dst_port | action |
|--------|---------------|---------------|----------|----------|--------|
| F1(1)  | 193.95.0.0/16 | 10.1.1.2      | 23       | tcp      | accept |
| F1(2)  | 193.95.0.0/16 | 10.0.0.0/8    | *        | *        | deny   |
| F1(3)  | 10.0.0.0/8    | 192.168.2.1   | 53       | udp      | accept |

|        | src_adr       | dst_adr        | protocol | dst_port | action |
|--------|---------------|----------------|----------|----------|--------|
| F2(1)  | 193.95.0.0/16 | 10.1.1.2       | 23       | tcp      | accept |
| F2(2)  | 10.0.0.0/8    | 192.168.2.2    | 22       | tcp      | deny   |
| F2(3)  | 193.95.0.0/16 | 192.168.2.0/24 | *        | *        | accept |
| F2(4)  | 10.0.0.0/8    | 192.168.2.1    | 53       | udp      | accept |

|        | src_adr       | dst_adr        | protocol | dst_port | action |
|--------|---------------|----------------|----------|----------|--------|
| F3(1)  | 193.95.0.0/16 | 192.168.2.0/24 | 22       | tcp      | deny   |
| F3(2)  | 193.95.0.0/16 | 10.1.1.2       | 23       | tcp      | accept |
| F3(3)  | 10.0.0.0/8    | 192.168.2.1    | 53       | udp      | accept |
| F3(4)  | 193.95.0.0/16 | 192.168.2.0/24 | *        | *        | accept |

Fig 9. A correct and complete distributed Configuration

## 5. Conclusion

In this paper, we propose a formal and automatic method for verifying that a distributed firewall configuration is conform to a security policy. Otherwise, the method provides key information helping users to correct configuration errors. Moreover, we also propose a procedure for checking and fixing the coherence of a security policy, which is a necessary condition for the conformance verification. Finally, our method has been implemented using the satisfiability solver modulo theories Yices. The experimental results obtained are very promising.

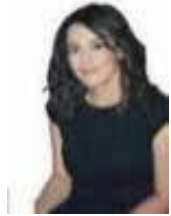

**Nihel Ben Youssef Ben Souayeh** received in 2007 his engineering degree in computer science from the National Institute of Applied Science and Technology. and she received in 2008 his MS degree from the Higer School of Communications of Tunis (Sup'Com). Nihel Ben Youssef Ben Souayeh received his Ph.D. degree in Information and Communication Technology from Sup'Com in 2012 and his research interests include network security, formal specification as well as formal validation and verification techniques. Nihel Ben Youssef Ben Souayeh works currently as assistant in the National Institute of Applied Science and Technology. She teaches computer security and software engineering. She is also member of Tunisian Association of Digital Security (TADS).

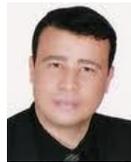

**Adel Bouhoula** obtained his undergraduate degree in computer engineering with distinction from the University of Tunis in Tunisia. He also holds a Masters, PhD and Habilitation from Henri Poincare University in Nancy, France. Adel Bouhoula is currently an Associate Professor at the SupCom Engineering School of Telecommunications in Tunisia. He is also the founder and Director of the Research Unit on Digital Security and the President of the Tunisian Association of Digital Security (TADS). His research interests include Automated Reasoning, Algebraic specifications, Rewriting, Network Security, Cryptography, and Validation of cryptographic protocols.